\documentclass[twocolumn,twocolappendix,trackchanges]{aastex63}
\usepackage{multirow,booktabs}
\usepackage{xcolor}
\received{March 22, 2020}
\revised{April 28, 2020}
\accepted{June 16, 2020}
\submitjournal{ApJ}

\shorttitle{T Tauri North}
\shortauthors{Flores et al.}
\graphicspath{{./}{figures/}}

\begin{document}

\title{Is T Tauri North a `Classical' T Tauri star?}

\email{caflores@hawaii.edu}
\correspondingauthor{Christian Flores}

\author[0000-0002-8591-472X]{C. Flores}
\author[0000-0001-8174-1932]{B. Reipurth}
\author[0000-0002-8293-1428]{M. S. Connelley}

\affiliation{Institute for Astronomy, University of Hawaii at Manoa, 640 N. Aohoku Place, Hilo, HI 96720, USA \\}




\begin{abstract}
We present high-resolution $H$- and $K$-band spectroscopic observations of the archetypal T~Tauri star T~Tau~North. Synthetic spectral modeling is used to derive the $K$-band temperature, surface gravity, magnetic field strength, and rotational velocity for this star. The $K$-band spectroscopic temperature measured is $\rm T_{K\text{-band}} = 3976 \pm 90 K$, which is $\sim 1000 \, \rm K$ cooler than the temperature measured from optical observations. Our $K$-band temperature measurement for T~Tau~N is confirmed using equivalent width line ratio vs. temperature relations in the $H$-band, from which a $\rm T_{H\text{-band}} = 4085 \pm 155 K$ is derived. This optical vs. IR temperature difference is interpreted as cool or hot spots, or both, covering a significant part of the surface of T~Tau~N. The gravity derived for T~Tau~N $\log{g}=3.45 \pm 0.14$ is lower than the gravity of nearly every other star in a sample of 24 classical T~Tauri stars in Taurus. Combining these temperature and gravity results with magnetic stellar evolutionary models, we find the age of T~Tau~N to be less than 1 Myr old. These results suggest that T~Tau~N is in an earlier evolutionary stage than most classical T~Tauri stars in Taurus, arguing that it is a protostar ejected from the embedded southern binary system shortly after its formation.
\end{abstract}



\section{Introduction} \label{sec:intro}

\cite{Joy1945} identified 11 irregular variable stars which seemed much alike, but different from other classes of variables, and named them after T~Tauri. In a now classic paper, \cite{Herbig1962} provided a detailed analysis of all aspects of the T~Tauri phenomenon.  While Herbig classified the T~Tauri stars in a restrictive manner, the term is used today in a more relaxed way, describing most {\em optically} visible young low-mass stars. The term young stellar object (YSO), coined by \cite{Strom1972}, tends to be used to describe all pre-main sequence objects, whether embedded or visible.  In more modern terms, the T~Tauri stars are generally equated with the Class~II objects (CTTS) and Class~III objects (WTTS).  Already by the middle of the 19th century, T~Tau attracted attention because of the presence of variable reflection nebulosity around it \citep{Hind1850,Barnard1895, Herbig1953}. The implication is that T~Tau is located very close to the surface of its small associated globule. For a historical account of the T~Tauri phenomenon, see \cite{Reipurth2016}.

Based on blue photographic spectra, \cite{Herbig1962} determined the spectral type of T~Tau to be dG5e. Already more than half a century ago this set T~Tau apart from most other T~Tauri stars, which almost as a rule are K- and M-stars. Since the work of Herbig, T~Tau has been observed at virtually all wavelengths and with all techniques. And in many cases, these observations set T~Tau apart as an atypical T~Tauri star.

\cite{Dyck1982} discovered an infrared companion to T~Tau only
0.7~arcsec to the south, which at a distance of 146.7$\pm$0.6~pc
\citep{Loinard2007} corresponds to a projected separation of
100~au. Subsequently, \cite{Koresko2000} resolved T~Tau~S into a close binary whose orbit by now has been sufficiently covered to provide a reliable orbit with a semi-major axis of 12.5~au. The more massive component of T~Tau~S has a mass of 2.12$\pm$0.10~M$_\odot$ and the companion has a mass of 0.53$\pm$0.06~M$_\odot$ \citep{Kohler2016}. 

In contrast to the southern components, the mass of T~Tau~N has only been calculated using optical spectral types combined with stellar evolutionary models, yielding mass estimates of 2.0~to~3.38~M$_\odot$ \citep{Koresko2000,White2004}. These estimates are, however, very sensitive to the adopted effective temperature of the star. Therefore, changes in the temperature of T~Tau~N might challenge current mass estimates for this star. This is the case, for example, for stars with large starspots on their surfaces such as LkCa~4 \citep{Gully-Santiago2017}, BP Tau \citep{Flores2019}, and possibly TW Hya \citep{Vacca2011}.

\cite{Bertout1983} suggested that T~Tau~S is a protostar, based on modeling of its radio spectrum. This is supported by analysis of the energy distribution of T~Tau~Sa, published by \cite{Ratzka2009}, which shows that it is a Class~I source. In a triple system as close as in the case of T~Tau, it is expected that the circumstellar or circumbinary disks are dynamically disturbed, a supposition that is borne out by high-resolution imaging and recent ALMA results \citep{Kasper2016, Yang2018,Long2019}. Another indication that the system is dynamically active is its association with the giant Herbig-Haro (HH) flow HH~355 \citep{Reipurth1997}, which is likely generated during close interactions in the newly formed T~Tau multiple system \citep{Reipurth2000}.

In this paper, we present iSHELL $H$- and $K$-band spectra of T~Tau~N from which stellar parameters such as temperature, gravity, and magnetic field strength are derived. The temperature and gravity measurements are combined with magnetic evolutionary models to estimate the age of T~Tau~N. Finally, T~Tau~N is put in context by comparing it to other classical T~Tauri stars in Taurus. In section \ref{sec:observations}, we detail the observations and data reduction. Section \ref{sec:modeling} summarizes the modeling technique used for the $K$-band data. In section \ref{sec:Results}, we present the stellar parameters obtained from the $K$-band spectrum in addition to temperature confirmation from the $H$-band spectrum. Section \ref{sec:Discussion} discusses possible ways of reconciling the optical and infrared temperature measurements as well as an explanation for the low surface gravity measured for T~Tau~N. In section \ref{sec:Conclusion} we present our conclusions.

\section{Observations and data reduction} \label{sec:observations}
We carried out observations with the IRTF 3.0-meter telescope on Maunakea, Hawaii, and used the high-resolution near-infrared echelle spectrograph iSHELL \citep{Rayner2016} to observe T~Tau~North on October 6, 2019. T~Tau~N was observed in the $H2$ mode i.e., from 1.55 to 1.74 $\mu$m and in the $K2$ mode i.e., from 2.09 to 2.38 $\mu$m; in both observations we used the 0\farcs75 slit to achieve a spectral resolution of R $\sim$ $50,000$. The observations were performed in excellent weather conditions with clear skies and a median seeing of 0\farcs7. Given that T~Tauri is a close triple system, with T~Tau~N being only $\sim$0\farcs7 North of T~Tau~S \citep[e.g.,][]{Csepany2015}, we guided on the brightest component (T~Tau~N) and placed the slit in the East-West direction to minimize the contamination from the southern binary source. T~Tau~S was not placed on the slit, but we estimate a 2-4\% $K$-band flux contamination from this component due to the observed seeing of 0\farcs7, a separation of 0\farcs7, and a $K$-band magnitude difference of $\Delta K \sim 2$  between T~Tau~N and T~Tau~Sa. This contamination is even smaller at $H$-band where the magnitude difference between the northern and southern component increases to $\Delta H \sim 6$ magnitudes \citep{Schaefer2014}.

Due to a known fringing issue in the iSHELL spectra, we decided to adopt the following observing strategy to improve the quality of the observations. We first guided on T~Tau~N and acquired five 200s $K2$ mode spectra of the star. We stopped guiding and immediately acquired a set of calibration spectra consisting of $K$-band flats and arc-lamp spectra. Then a nearby telluric A0 standard star was observed within 0.1 airmasses of T~Tau and integrated on it long enough to achieve five $S/N\sim120$ spectra. Afterward, the guiding was stopped again and a second set of $K$-band calibration files was obtained to correct the telluric standard star. The same process was repeated to acquire $H2$ mode spectra of T~Tau~N. At the end of the observing night, dark frames that matched the integration times of our science target and telluric standard stars were obtained. 

The data were reduced using the IDL-based package \texttt{Spextool v5.0.2}\footnote{\url{http://irtfweb.ifa.hawaii.edu/research/dr_resources/}} \citep{Cushing2004}. First,  normalized flat field images and wavelength calibration files were created using the \texttt{xspextool} task. Then, the spectra of each star (T~Tau and telluric standards) were extracted using the point source extraction configuration. To increase the S/N of the final spectra, we used \texttt{xcombspec} to combine the individual multi-order stellar spectra and \texttt{xmergeorders} to merge multi-order spectra into a single continuous spectrum. Afterward, \texttt{xtellcor} task \citep{vacca2003} was used to remove atmospheric absorption lines from the science spectra, and \texttt{xcleanspec} was used to eliminate any deviant or negative pixels caused by imperfections in the infrared array. Finally, we continuum-normalized both the $H2$ and $K2$ reduced spectra using a first order broken power-law function with the  \texttt{specnorm}\footnote{\url{ftp://ftp.ster.kuleuven.be/dist/pierre/Mike/IvSPythonDoc/plotting/specnorm.html}} code. 

\section{Stellar parameter measurements}
\label{sec:modeling}
To derive stellar parameters for T~Tau~N, we used a synthetic stellar spectrum technique that combines MARCS stellar atmospheric models \citep{Gustafsson2008}, the polarized radiative transfer code MoogStokes \citep{Deen2013}, along with atomic and molecular lines from the VALD3 \citep{Pakhomov2019,Ryabchikova2015} and HITEMP databases \citep{Rothman2010}. Details of the modeling technique can be found in \cite{Flores2019}. Here only the important parts of the method are summarized as well as additional steps included for this particular study.

\begin{deluxetable}{ccc}
\tablecaption{Spectral Regions Used in the Computation of the Synthetic Models. \label{table:wavelength_regions} }
\tablehead{
\colhead{Region number} & \colhead{Spectral range}  & \colhead{Main spectral} \\ 
  & (\micron)  & lines }
\startdata
 1 & 2.1091--2.1107 & Al, Fe  \\
 2 & 2.1163--2.1176 & Al, Fe \\
 3\tablenotemark{a} & 2.1780--2.1792 & Si, Ti \\
 4\tablenotemark{a} & 2.1896--2.1908 & Fe, Ti \\
 5 & 2.2055--2.2100 & Na, Sc, Si \\
 6 & 2.2200--2.2336 &  Ti, Fe, Sc \\
 7 & 2.2608--2.2664 & Ca, Fe, Ti, \\
 8 & 2.2930--2.3042 & CO(J=2-0) \\
\enddata
\tablenotetext{a}{Default \texttt{VALD3} oscillator strengths and van der Waals constant values were modified for lines in these regions (see Appendix \ref{section:vald3_modification}).}
\tablecomments{Wavelength values are defined in vacuum.}
\end{deluxetable}

Due to possible contamination of  the stellar spectra from  disk emission, a two step approach to derive the stellar parameters of T~Tau~N was performed. In step one, the spectrum of T~Tau~N was modeled varying seven stellar parameters: temperature, gravity, magnetic field strength, projected rotational velocity, $K$-band veiling, micro-turbulence, and CO abundance. In this first approach all eight wavelength regions defined in Table \ref{table:wavelength_regions} were used. This first calculation, which includes the CO lines, allows us to estimate the projected rotational velocity of T~Tau~N, as the CO lines are known to be insensitive to the Zeeman effect and are therefore an excellent tracer of the rotational broadening of the star. As a second step, the rotational velocity of the star was fixed to the value obtained in step one and the CO region was ignored. Now only five of the stellar parameters were varied: temperature, gravity, magnetic field strength, $K$-band veiling, and micro-turbulence. By doing this we attempt to avoid contamination from disk emission in the CO region that could affect temperature and/or gravity measurements of the star \citep[e.g.,][]{Doppmann2005}.

In our models, we adopted a solar metallicity, a single magnetic field strength component to characterize the average magnetic field over the visible stellar surface, and we used an IR $K$-band veiling that is independent of wavelength. The 7-D (5-D) parameter space of our models was explored with a Markov Chain Monte Carlo (MCMC) method using the \texttt{emcee} package \citep{Foreman-Mackey2013}. In this framework, a log-likelihood metric was maximized assigning an uninformed prior on each of the stellar parameters. Our runs started with 50 individual chains (walkers) advancing a total of 1500 steps. Convergence of the chain was assessed by examining the time-line plot and discarding the first 300 steps from each chain as a burn-in stage. The remaining 50,000 models were kept to derive the best stellar parameters as the median of the distribution and the formal errors of our MCMC chain are reported as the 16th and 84th percentile of the distributions.

\section{Results}
\label{sec:Results}

\subsection{Stellar parameters of T~Tau~N from K-band}
\label{subsection:stellar-params-K-band}

\begin{deluxetable}{lccc}
\tablecaption{Stellar parameters of T Tau N. \label{table:Stellar_params}}
\tablehead{ & \colhead{$K$-band} & \colhead{$H$-band} & \colhead{Optical-bands}\tablenotemark{a}}
\startdata
$\rm Temperature$ (K) & $3976^{+90}_{-90}$ & $4085^{+155}_{-155}$  & 4870    \\
$\log{g}$ & $3.45^{+0.14}_{-0.14}$ & - & - \\
$\langle \rm B \rangle $ (kG) & $1.99^{+0.04}_{-0.05}$ & - & - \\
$v\sin(i)$ (km s$^{-1}$) & $20.8^{+0.17}_{-0.13}$ & - & -\\
veiling & $3.0^{+0.04}_{-0.04}$ & - & 0.1 \\
$v_{micro}$ (km s$^{-1}$) & $0.19^{+0.11}_{-0.06}$ & - & -\\
S/N & 281 & 248 & ? \\
\enddata
\tablenotetext{a}{Results from \cite{Herczeg2014}}
\tablecomments{The signal-to-noise ratio corresponds to the  reported median S/N over the full $H2$- and $K2$-mode spectra from \texttt{Spextool}.}
\end{deluxetable}

\begin{figure*}[ht]
\epsscale{1.0}
\plotone{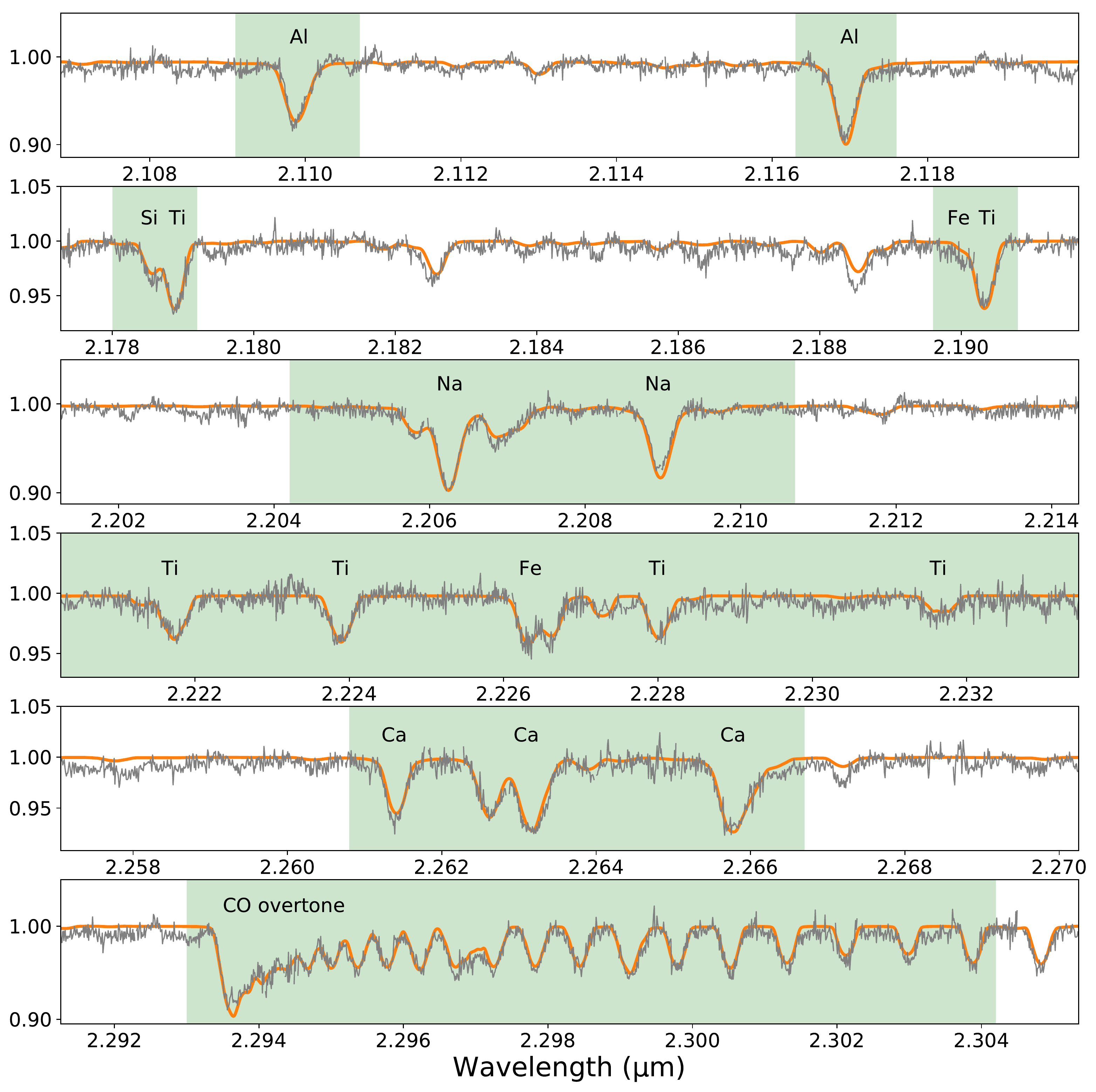}
\caption{Comparison between the observed spectrum of T~Tauri~N (in gray) and our best-fit model (in orange) from MoogStokes. The eight green shaded regions show the wavelength regions used to fit the spectrum of the young star in our first iteration. In our second iteration we used all the regions but the CO overtone (see Section \ref{sec:modeling}). \label{fig:best_fit}}
\end{figure*}

The spectrum of T~Tau~N was modeled using the formalism defined in Section \ref{sec:modeling}. In the first iteration, when the CO region was included, we obtained a temperature of $\rm T_{K\text{-band}} = 4050^{+91}_{-93}$ K, a surface gravity of $\log{g}=3.70^{+0.15}_{-0.15}$, a magnetic field strength of $\rm \langle B \rangle = 2.25^{+0.08}_{-0.08}$ kG, a $K$-band veiling of $r_K=3.26^{+0.06}_{-0.06}$, a $v_{micro}=2.4^{+0.30}_{-0.42}\rm \, km\,s^{-1}$, and a projected rotational velocity of $v\sin{i} = 20.8^{+0.17}_{-0.13} \, \rm km\,s^{-1}$. We noticed that the strength of the CO absorption was significantly weaker than what is expected from the derived stellar parameters alone. This weaker CO absorption was compensated in our models by the CO abundance parameter, which reduced the CO depth by 2$\%$ to better match the data. The shape of the CO overtone was also distorted with respect to the models, which indicates possible disk emission. These features are further discussed in Subsection \ref{subsec:disk_emission}. For these reasons and as described in the previous section, the CO region was subsequently excluded and the projected rotational velocity was fixed to the best fit value found before. In the second iteration, a temperature of $\rm T_{K\text{-band}} = 3976^{+90}_{-90}$ K, a surface gravity of $\log{g}=3.45^{+0.14}_{-0.14}$, a surface magnetic field strength of $\rm \langle B \rangle = 1.9^{+0.04}_{-0.05}$~kG, a $K$-band veiling of $r_K=3.0^{+0.04}_{-0.04}$, and a $v_{micro}=0.19^{+0.11}_{-0.06} \rm \, km\,s^{-1}$ were measured. As explained in Section \ref{sec:modeling}, we trust that the second set of values better represent the stellar parameters of T~Tau~N as they are not affected by a contaminated CO region. This second set of results is summarized in the first column of Table \ref{table:Stellar_params}. The uncertainties quoted in column 1 correspond to the sum in quadrature of the formal uncertainty from the MCMC models and the empirical uncertainties calculated in \cite{Flores2019}. In Figure \ref{fig:best_fit}, the $K$-band spectrum of T~Tau~N is plotted in gray and the MoogStokes model with the best model parameters from Table \ref{table:Stellar_params} is overplotted in orange.

Although we acquired a high S/N spectrum of T~Tau~N in the $K$-band (S/N$\sim281$), one sees from Fig \ref{fig:best_fit} that the absorption lines are only 10-20x stronger than the noise in the spectrum. For example the Na line at 2.206 $\mu$m, which is one of the strongest lines in the spectrum of K- and M-type stars, only reached $10\%$ of the depth of the continuum in the spectrum of T~Tau~N. This reduced depth of the absorption lines is due to the IR veiling, which is strong in the spectrum of T~Tau~N with a value of $r_{K}=3.0$. This excess of $K$-band continuum over the photospheric stellar emission is likely produced by hot dust at the inner rim of the circumstellar disk of T~Tauri~N \citep{Folha1999, Johns-Krull2001}.

\begin{figure*}[ht]
\epsscale{1.1}
\plotone{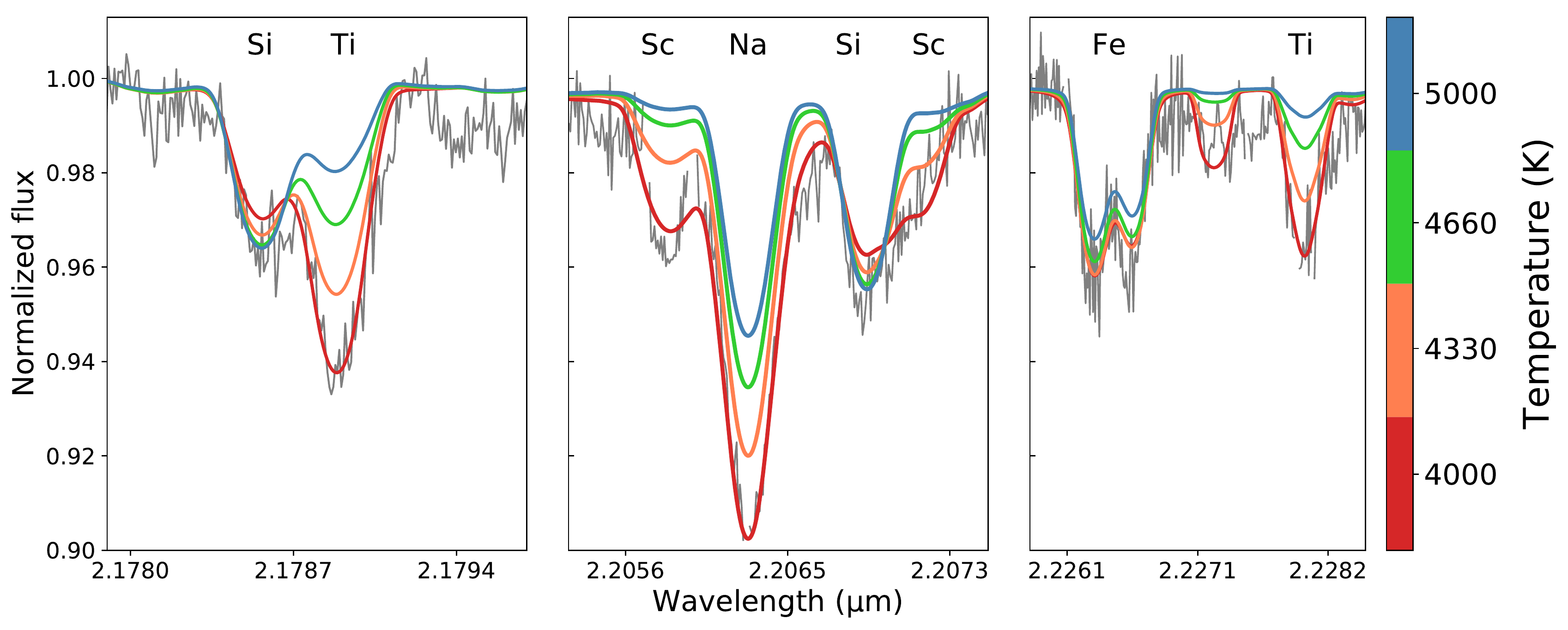}
\caption{Comparison between MoogStokes models with different temperatures (color lines) vs. the spectrum of T~Tau~N (grey line) in three wavelength regions. We see how the line-depth ratio of the Si/Ti (left), Sc/Na (middle), and Fe/Ti (right) substantially change as the temperature of the models vary. The best match in all cases corresponds to a model with a temperature of $\sim$ 4000 K. \label{fig:temp_change}}
\end{figure*}

Since optical determinations of the spectral type of T~Tau~N typically show it as a K0 star \citep{Herbig1977,White2004} with a temperature of $\sim$5000~K depending on the spectral type to temperature conversions adopted \citep{Herczeg2014}, we tested how models with different temperatures fit the $K$-band spectrum of T~Tau~N. Figure \ref{fig:temp_change} shows three different parts of the $K$-band spectrum of T~Tau~N which have strong sensitivity to variations in temperature in the range of 4000~K to 5000~K. For each of the regions shown in Figure \ref{fig:temp_change}, MoogStokes models were computed using our best fit parameters (see Table \ref{table:Stellar_params}), but with temperatures that span the range of values between our best fit temperature ($\rm T_{K\text{-band}}\sim4000$~K) and optical measurements ($\rm T_{opt}\sim5000$~K). One sees from Figure \ref{fig:temp_change} that stellar models with $\rm T_{eff}>4000 \,K$ fail to reproduce the line depth ratios of the observations. This is better seen in lines with opposite temperature sensitivity such as the Si + Ti or the Si + Sc lines of Figure \ref{fig:temp_change}, which due to their temperature behavior become powerful temperature diagnostics \citep{Doppmann2003}. We also note that such changes in line depth ratios cannot be explained by variations of the IR veiling, rotational velocity, or micro-turbulence as these processes affect all the photospheric lines in similar ways. The fact that a $\sim$4000~K temperature model fits the $K$-band spectra of T~Tau~N does not imply that previous optical measurement are erroneous. It might rather suggest the presence of inhomogeneities on the surface of T~Tau~N. In Subsection \ref{subsec:optical_vs_NIR_measurements}, we discuss possible ways of reconciling both optical and near-infrared observations.

\begin{figure}[ht!]
\epsscale{1.25}
\plotone{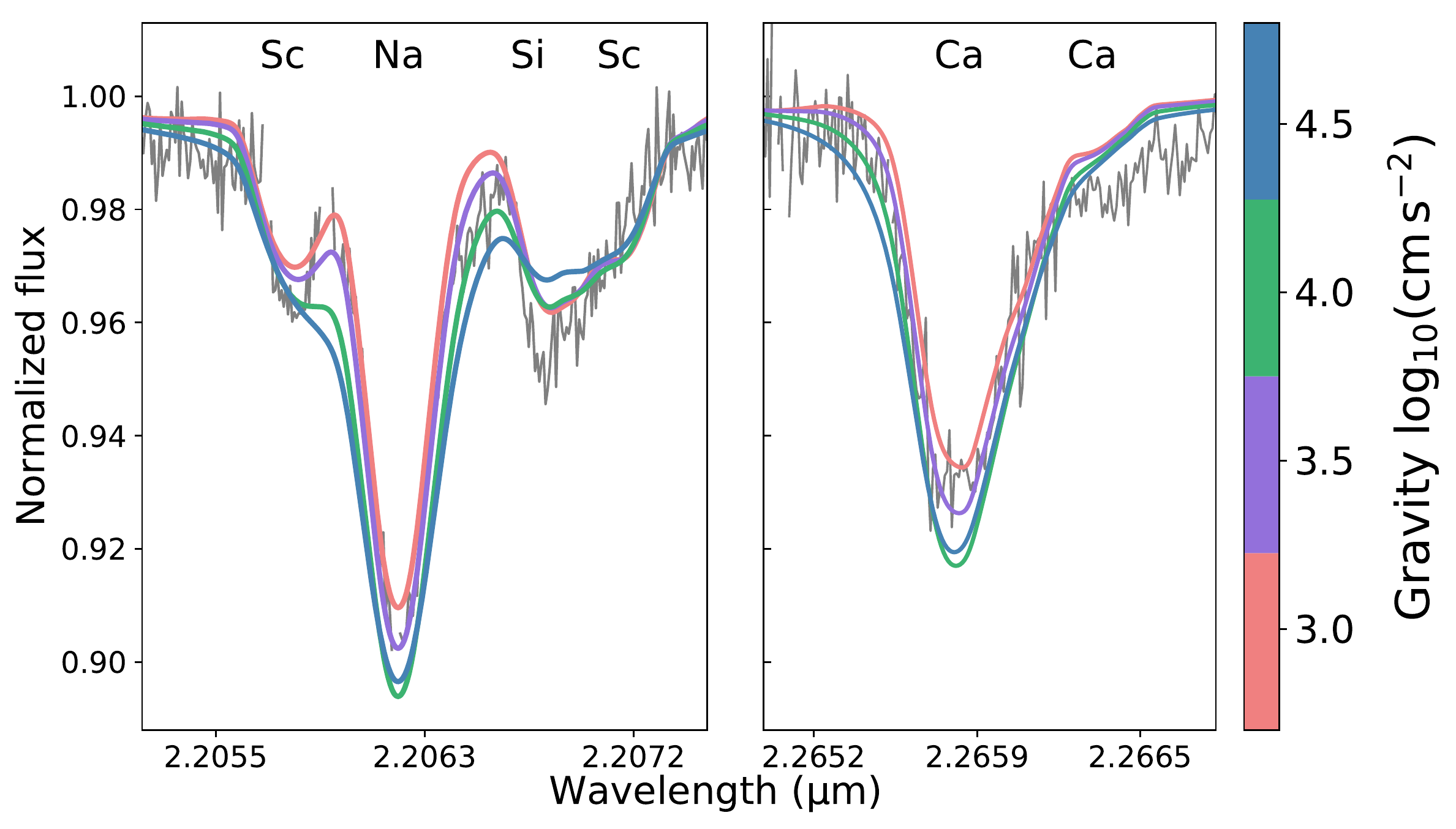}
\caption{Comparison between MoogStokes models with different gravities vs. the spectrum of T~Tau~N in the Na 2.206 $\mu$m region (left) and the Ca 2.265 $\mu$m region (right). Models with gravities larger than $\log{g}>3.5$ overestimate the broadening of the Na and Ca lines, particularly where the Na and the Sc lines meet. A model with $\log{g}=3.0$ better reproduces the gap between the Na and Sc lines, but it fails to reproduce the width of the Ca line. \label{fig:grav_change}}
\end{figure}

We now turn to the effects that the surface gravity parameter has on the stellar models. Figure \ref{fig:grav_change} shows two wavelength regions where the synthetic models display the largest changes with the surface gravity parameter. In these regions, MoogStokes models were computed using our best fit stellar parameters, except for the surface gravity which was varied from 3.0 to 4.5 in log scale of $\rm cm \, s^{-2}$. We note that changes in surface gravity produce milder effects in the model spectra than changes in effective temperature (see Figure \ref{fig:temp_change}), however, gravity variations are easily recognizable due to the broadening effect they imprint in the spectra. Figure \ref{fig:grav_change} shows that as the gravity of the star increases, the pressure broadening affects the Na line and makes it blend with the Sc line next to it (left panel). The pressure broadening also affects the Ca line (right panel) causing the models to overestimate the width of the line for gravities higher than $\log{g} \sim 3.5$. An interesting point to note is that a surface gravity of $\log{g} \sim 3.0$ seems to better fit the left wing of the Na line, but it fails to reproduce both the right wing of the Na line (left panel) and the width of the Ca line (right panel). Our best fit model yields a surface gravity of $\log{g} = 3.45 \pm 0.14$, which is likely a compromise between these different effects. In Appendix \ref{section:appendix_stellar_params}, we additionally show that the gravity value presented here is broadly consistent with what we expect for the rotational velocity (see below), period measurement, and stellar inclination angle of T~Tau~N.

A magnetic field strength of $\langle \rm B \rangle = 1.99^{+0.2}_{-0.2}$ kG is measured for T~Tau~N; which is consistent with results from \cite{Johns-Krull2007} who measured a $\rm \langle B \rangle = 2.37 \pm 0.3$ kG. Since \cite{Johns-Krull2007} obtained observations of T~Tau~N over 20 years ago, it is difficult to assess whether the (statistically insignificant) difference in magnetic field strengths is due to variability or if it is due to differences in the modeling techniques. \cite{Johns-Krull2001} and \cite{Folha1999} reported infrared $K$-band veiling measurements of $\rm r_k=2.74 \pm 0.29$ and $\rm r_k=2.5 \pm 0.7$, respectively, both being consistent with our $\rm r_k$ measurements. Likewise, our projected rotational value derived for T~Tau~N of $v\sin{i}= 20.8^{+0.46}_{-0.38}$ $\rm km \, s^{-1}$, is consistent with values from \cite{Hartmann1989}, \cite{Johns-Krull2001}, and \cite{White2004}.

This latter result confirms that most of the NIR absorption lines in the spectrum of T~Tau~N have a stellar origin. Otherwise, the observed absorption lines would have formed, for an assumed stellar mass of 1$\rm M_\odot$  (2$\rm M_\odot$), at a distance of 2.1 au (4.2 au) away from the star in order to match the rotational velocity of $v\sin{i}=20.8 \,\rm km\, s^{-1}$. Although it is not impossible to produce a $\sim4000$~K emission at those distances, the typical distance to the inner rim of a disk is only a few stellar radii i.e, 0.03 to 0.06 au from the star.

\subsection{Temperature measurement from the H-band spectrum}
\label{subsect:H-band_temp}


To confirm the temperature measurement obtained for T~Tau~N in the previous section, we now use empirical temperature relations in the $H$-band. \cite{Lopez-Valdivia2019} used IGRINS \citep{Park2014} $H$-band spectra to construct a correlation between line-depth ratios of atomic and molecular lines to spectral temperatures for main-sequence stars in the temperature range of $\sim$2800~K to $\sim$4500~K. They found that the OH line at 1.5620~$\mu$m and the Fe line at 1.5630~$\mu$m are excellent temperature indicators for stars hotter than 3200~K. Although this technique works best for main-sequence stars, \cite{Lopez-Valdivia2019} explored how different metallicities, rotation rates, and gravities could influence the temperature results. For the whole sample of stars, ($\rm T_{eff} \sim$2800~K to $\sim$4500~K) \cite{Lopez-Valdivia2019} reported a `cumulative uncertainty budget' of 140~K in terms of calibration errors, while systematic errors reached differences as high as 120~K. However, for stars hotter than $\rm T_{eff}=4100$~K, they found that gravity and rotational velocity variations produced temperature differences of the order of $\sim$20~K; and changes in metallicity of $\pm$0.4~dex produced variations in temperature of $\pm 100$~K.

\subsubsection{Error Analysis}
We carefully continuum-normalized the $H$-band spectrum of T~Tau~N following the method described in \cite{Lopez-Valdivia2019}. We measured the equivalent width (EW) of the OH 1.5620~$\mu$m and the Fe 1.5630~$\mu$m lines using a Simpson integration algorithm, then we divided them to obtain a Fe/OH EW line ratio. To understand the uncertainties in the measured EW line-depth ratio, we took two different approaches. In the first approach, three different continuum normalization levels were chosen: one close to the highest peak of the spectrum, another going through the median of the pseudo-continuum, and finally one just below the pseudo-continuum. All of them are acceptable eyeball normalization options (see Fig. \ref{fig:H-band-temperature}). In each case, the EW ratios of the Fe and OH lines were calculated and then the Fe/OH EW line ratios were computed. The three values of the EW line ratios were averaged to obtain a $ \rm \overline{Fe/OH}=2.61$ with a standard deviation of $\Delta \rm Fe/OH=0.13$. The second approach was to perform the error propagation of the Simpson integration for the EW measurements of the Fe and OH lines, and then further propagate the uncertainties in the division of both quantities. In this case, a formal error of $\Delta \rm Fe/OH_{formal} =0.08$ was obtained, which is smaller than the uncertainty obtained for different continuum normalization levels. Therefore, the formal and normalization uncertainties were added in quadrature to obtain a total uncertainty of $\Delta \rm Fe/OH_{total} =0.15$ in the EW line-depth ratio measurements.

\subsubsection{Temperature measurement}
\label{subsect:H-band_temp_meas}
\begin{figure}[ht!]
\epsscale{1.1}
\plotone{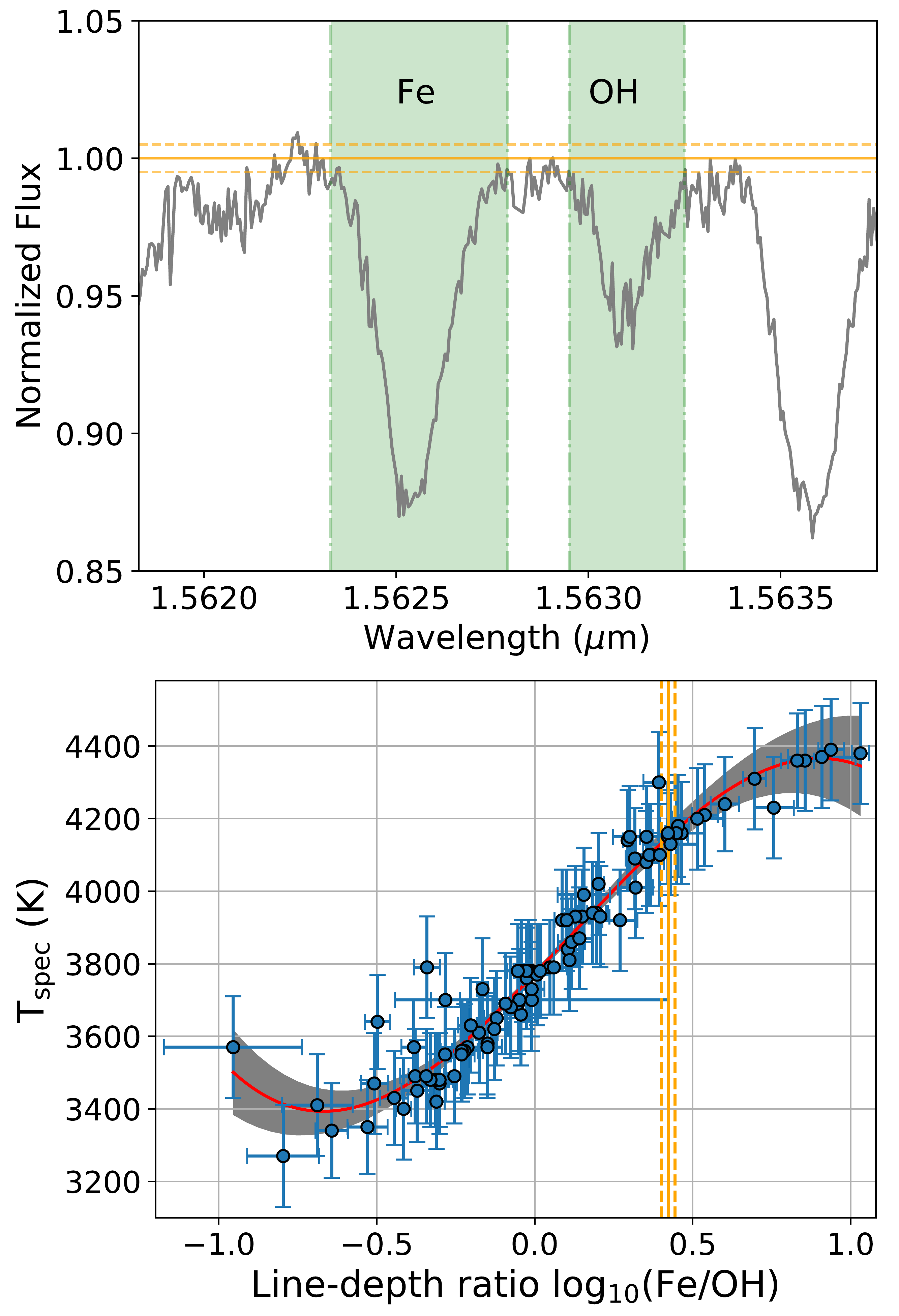}
\caption{Top panel: We show a portion of the $H$-band spectrum of T~Tau~N centered on the Fe/OH region at 1.563 $\mu$m. We highlight the Fe 1.5625 $\mu$m and the OH 1.530 $\mu$m lines with green colors in the region where their EWs were calculated. The three continuum normalization levels used to compute the line-depth ratio uncertainties are marked as golden dashed lines centered around 1.0. Bottom panel: We reproduced the Fe/OH EW line-depth ratio v.s. spectral temperature plot from \cite{Lopez-Valdivia2019} using 252 M and K dwarfs from their sample. The measurements for the dwarfs are shown as blue circles, the red line corresponds to a 3rd order polynomial fit in log(Fe/OH) space to the data, and the grey region is the uncertainty in the polynomial fit. The golden vertical lines correspond to the log(Fe/OH) value measured for T~Tau~N (solid) and its measured total uncertainty (dashed). \label{fig:H-band-temperature}}
\end{figure}

Given that the linear relationship between the line-depth ratio of Fe/OH and temperature presented in \cite{Lopez-Valdivia2019} works better in the temperature range of $\sim$3000~K to $\sim$3900~K, we have replotted the data from \cite{Lopez-Valdivia2019} in such a way that the temperature measured for T~Tau~N could be better visualized. The top panel of Figure \ref{fig:H-band-temperature} shows a small part of the spectrum of T~Tau~N centered around the OH 1.5620 $\mu$m and the Fe 1.5630 $\mu$m lines, marked with green dashed regions. In the bottom panel of Figure \ref{fig:H-band-temperature}, we show the replotted EW line-depth ratio of Fe/OH vs. $\rm T_{spec}$ using 252 of the sources with Fe and OH measurements from \cite{Lopez-Valdivia2019}. The line-depth ratio was plotted on a logarithmic scale and the spectroscopic temperature on a linear scale. A third order polynomial curve was fit to the data using the \texttt{polyfit} python package from which we obtained best fit coefficients $f(x)= - 501\pm 56\, x^3 + 175\pm34\, x^2 + 908\pm31\, x + 3772\pm8$, with $x$ being the $\log_{10}$ of the EW line depth ratio between Fe and OH. The calculated uncertainty in the polynomial coefficients was obtained by computing the square root of the diagonal of the covariance matrix. Using the above mentioned formula with the EW line-depth ratio measured for T~Tau~N, a spectroscopic temperature of $\rm T_{H\text{-band}}=4145 \pm 40$~K was obtained. Given the uncertainties in the line-depth ratio method \cite[see Section 4.1 in][]{Lopez-Valdivia2019} and following priv. communication with R. L\'opez-Valdivia, we included an additional temperature uncertainty of 150~K and a 60~K correction to the value obtained due to systematic effects. Folding everything together, we derived a corrected spectroscopic $H$-band temperature of $\rm T_{H\text{-band}}=4085 \pm 155$~K for T~Tau~N.

\section{Discussion}
\label{sec:Discussion}

\subsection{Disk CO emission in the spectrum of T~Tau~N}
\label{subsec:disk_emission}

\begin{figure*}[ht]
\epsscale{1.1}
\plotone{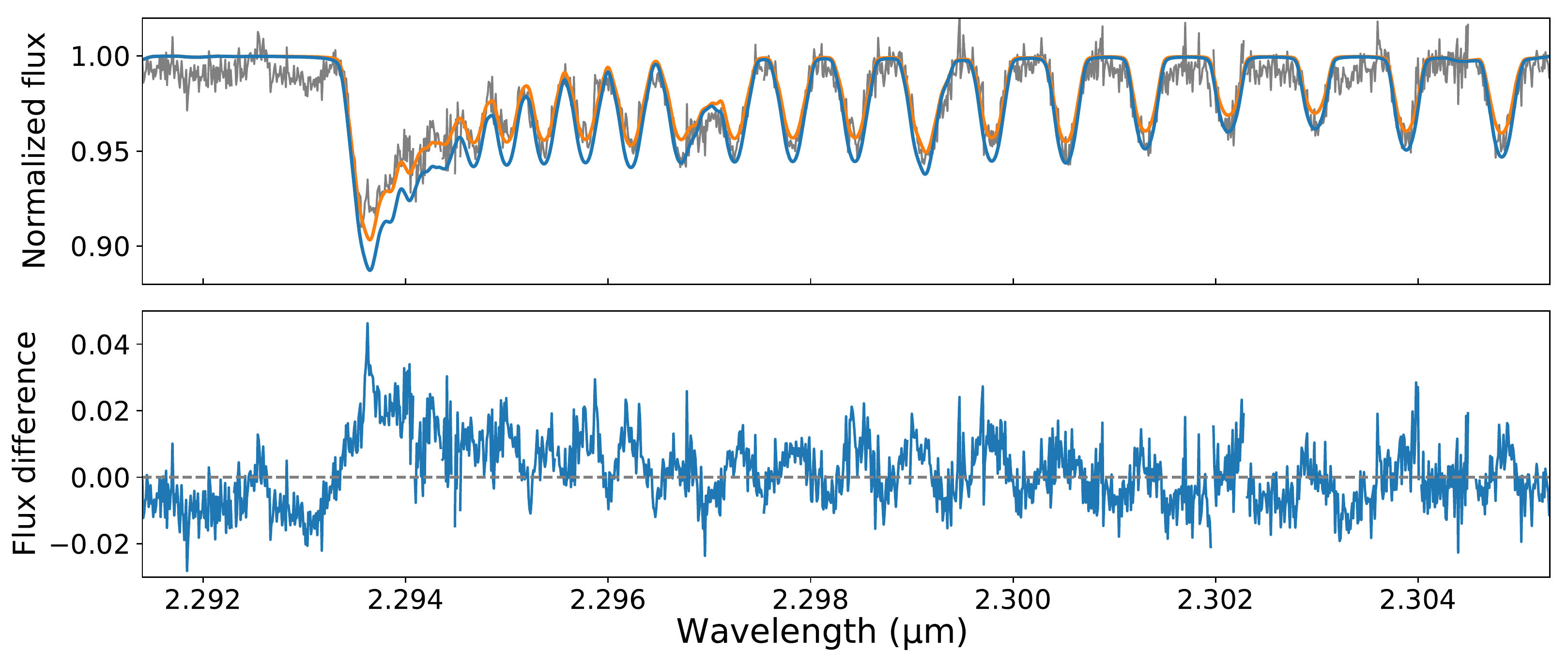}
\caption{Top panel shows the spectrum of T~Tau~N centered on the CO region (in gray), the best fit model with reduced CO abundance (in orange), and the best fit model with Solar CO abundance (in blue). The bottom panel shows the difference between the model with Solar CO abundance and the spectrum of T~Tau~N. A dashed gray line marks the zero flux difference for illustrative purposes. Notice that the y-scale in the bottom panel is different than in the top panel.\label{fig:contaminated_CO}}
\end{figure*}

In Section \ref{sec:Results}, we suggested that the strength of the CO absorption lines in the spectrum of T~Tau~N were weaker and also seemed distorted when compared to synthetic stellar models. A zoomed-in version of the bottom panel of Figure \ref{fig:best_fit} is presented in the upper panel of Figure \ref{fig:contaminated_CO}. The spectrum in orange corresponds to the best fit model of T~Tau~N with reduced CO abundance, the grey spectrum is the data, and the blue line represents the best fit model spectrum of T~Tau~N, but with a solar CO abundance value. Both synthetic models only differ by a scaling factor. Although the orange model (reduced CO abundance) better fits the depth of the CO bandhead, it underestimates the depth of most of the CO lines redward of 2.299~$\micron$. The opposite is true for the blue model (solar CO abundance), which overestimates the depth of the CO bandhead but fits  better the CO lines on the right side of the plot. The bottom panel of Figure \ref{fig:contaminated_CO} shows the residual between the data and the model with solar CO abundance, which displays an emission-like spectrum. A strong CO emission at the bandhead and a declining flux towards the `bandtail' are features typically associated with a CO disk emission spectrum  in young stars  \citep{Carr1993,Carr2004,Najita1996}. The residual flux detected in T~Tau~N, although very weak in strength ($\sim2\%$ of the depth of the continuum), is likely to represent the disk's contribution to the stellar absorption spectrum of the star. Cases like this, where the observed spectrum is a combination of flux from the stellar photosphere and flux from the disk have been reported in the literature, such as the case of GV~Tau~S and GV~Tau~N observed and modeled by \cite{Doppmann2008} using NIRSPEC $K$-band data.  

As a consequence of this residual emission seen in the spectrum of T~Tau~N, we justify the exclusion of the CO region (see Section \ref{sec:modeling}) when deriving the stellar parameters of the star. 

\subsection{T Tauri~N compared to other CTTS in Taurus}

To provide a direct comparison between T~Tau~N and other CTTS in Taurus, we have modeled the $K$-band spectra of a sample of 24 classical T~Tauri stars using the same formalism described in Section \ref{sec:modeling} (Flores et al., in prep). Here only the $\rm T_{K\text{-band}}$ and $\log{g}$ values are presented. Comparing the stellar properties of T~Tau~N with the Taurus sample, we see that the gravity derived for T~Tau~N of $\log{g}=3.45\pm0.14$ is the second lowest gravity and it is significantly lower than the median gravity measured for the 24 CTTS of $\langle \log{g} \rangle = 3.8 \pm 0.3$. T~Tau~N is among the hottest stars of the sample with a $K$-band temperature of $\rm T_{K\text{-band}} = 3976 \pm 90$ K, while the median temperature derived for the CTTS in Taurus is $\langle T_{K\text{-band}} \rangle =3514\pm 193$~K .

\begin{figure*}[ht!]
\epsscale{1.1}
\plotone{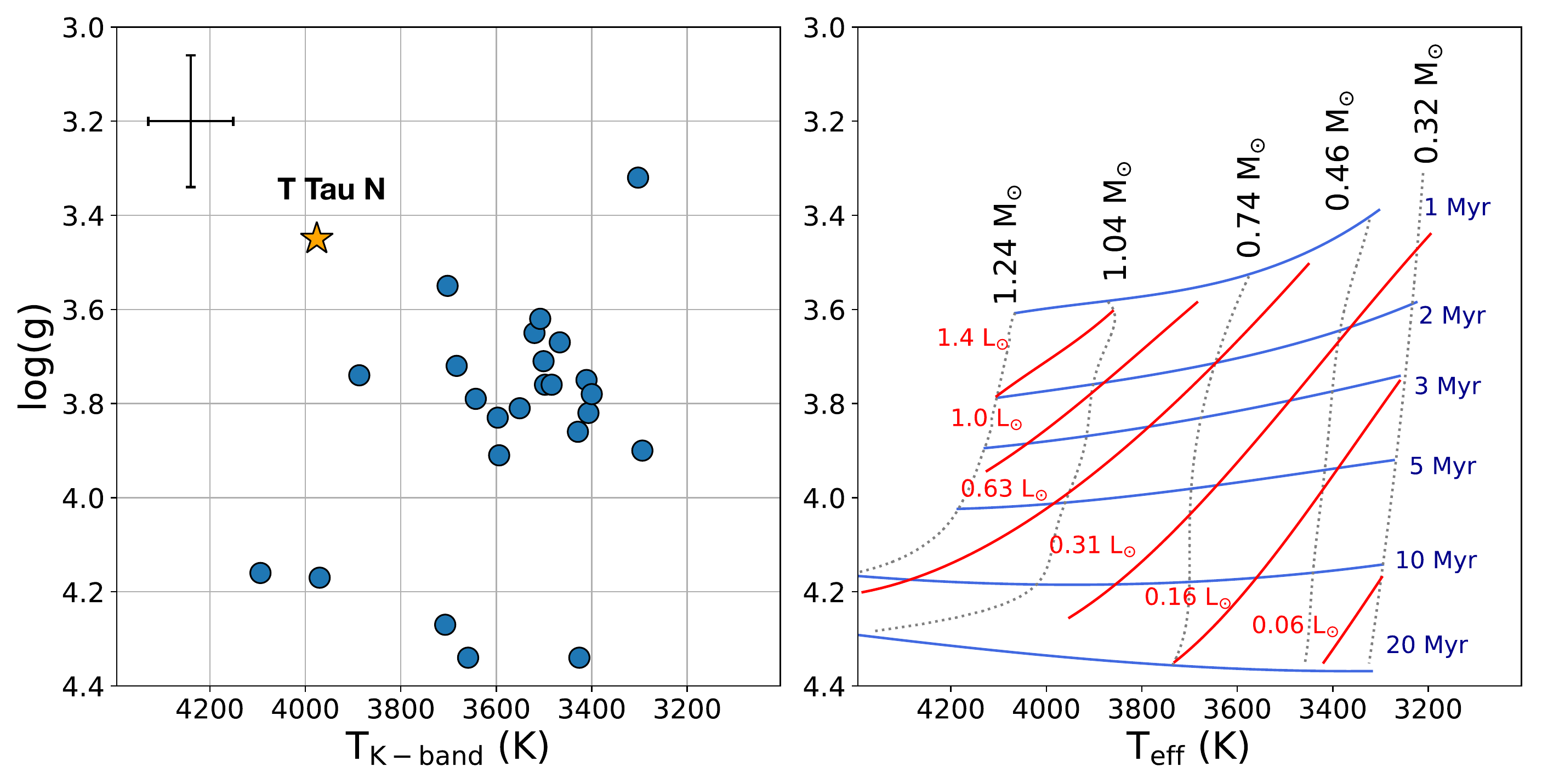}
\caption{The left panel shows the position of T~Tau~N (yellow star) and a sample of 24 CTTS from Taurus (blue circles) in the $K$-band temperature vs. surface gravity diagram. A sample-representative uncertainty of 90~K in $K$-band temperature and 0.14~dex in gravity is shown in the upper-left part of the plot. The right panel show isochrones (solid blue), evolutionary tracks (dotted black), and luminosity curves (solid red) from the \cite{Feiden2016} magnetic evolutionary models.  \label{fig:evol_models}}
\end{figure*}

The above mentioned results can be visualized on the left panel of Figure \ref{fig:evol_models}, where the $K$-band temperature and the gravity of the 25 stars are displayed. On the right panel of this figure,  mass tracks, isochrones, and stellar luminosities from the \cite{Feiden2016} magnetic evolutionary models are shown. It is worth noting that the y-axes in both panels of Figure \ref{fig:evol_models} are the same ($\log{g}$), but the x-axes are different; our measurements correspond to $K$-band temperatures (left panel) while the evolutionary models (shown on the right panel) are calculated using the effective temperature of the stars. These two temperatures coincide for stars with thermally homogeneous photospheres (non-spotted stars), however, such condition is unlikely to be met by our young and magnetically active stars.

Given that the isochrones shown on the right panel of Figure \ref{fig:evol_models} are almost horizontal in the $\log{g}$ vs. temperature space (i.e., they are highly insensitive to temperature errors), we find that most CTTS in Taurus (18) have gravities consistent with ages of 1 to 5 Myr old, a few of them (5) have gravities that suggest ages of 10 to 20 Myr old, but only T~Tau~N and one other star have gravities consistent with ages of less than 1 Myr old. The mass tracks, on the other hand, are almost entirely defined by the effective temperature of the stars (at least in the mass ranges shown in the plot and for ages $<10$~Myr), which implies that deriving masses using the $K$-band temperatures might lead to incorrect results.

Finally, we note that since stellar effective temperatures have not been measured in this work, comparing luminosities between stars is only meaningful in a relative sense and not necessarily in terms of absolute luminosity values.  That being said, we find that T~Tau~N  occupies a place in the temperature vs. gravity space that makes it the most luminous star of all our sample, with a luminosity of 2.6 times the luminosity of the second most luminous star in this work, and more than seven times the median luminosity of the sample. This result is also supported by photometric data which show that T~Tau~N is one of the most luminous T~Tauri stars known.

\subsection{Optical vs. NIR temperature measurements}
\label{subsec:optical_vs_NIR_measurements}
The archetypal T~Tauri star T~Tau has been studied for over 70 years; during this time many studies have spectral typed the northern (optically brightest) component of the system. T~Tau~N was classified as a dG5e type by \cite{Joy1945} and \cite{Herbig1962} using spectrograms acquired on photographic plates. Later on, \cite{Herbig1977} re-classified T~Tau as a K0eIV star using an optical $\lambda = 5850$~\AA \, to $\lambda =6700$~\AA \, spectrogram taken with the Coud\'e Spectrograph at the Lick $120''$ telescope at a spectral resolution of 34~\AA \, $\rm mm^{-1}$. More recent spectral type measurements of T~Tau~N have yielded similar results. \cite{White2004} measured a spectral type of $\rm K0 \pm 2 $ using a $\lambda =  \rm 6330$~\AA \, to $\lambda =8750 $~\AA \, Keck HIRES spectrum at a resolution of $\rm R \sim 37,000$. \cite{Herczeg2014} derived a $\rm K0 \pm 1$ spectral type for T~Tau, using the R5150 spectral index (produced by MgH absorption) with data acquired from the Double Spectrograph on the Hale 200 inch Telescope at Palomar with a spectral resolution of $\rm R \sim 700$. 

The associated effective temperature of a K0 spectral type greatly depends on the conversion scale used. Older spectral type to temperature conversion scales developed for field dwarfs assigned a $\rm T_{eff} =5236$ K to a K0 type \citep{Cohen1979}, while \cite{Kenyon1995} used a  $\rm T_{eff} =5250$ K for a K0 spectral type star, and \cite{Herczeg2014} re-defined the K0 spectral type to $\rm T_{eff} =4870$ K by fitting Phoenix/BT-Settl synthetic spectra to a grid of young spectral templates (K5-M8.5) and a sample of F-K3 luminosity IV stars from the Pickles library \citep{Pickles1998}. Regardless of the spectral type to temperature conversion used, our $K$-band and $H$-band temperature measurements are significantly cooler than optical temperature measurements (see Table \ref{table:Stellar_params}), with an optical to NIR temperature difference of $\Delta T \sim 800 - 1200$~K. 

This temperature difference is larger than similar measurements in other T~Tauri stars. For example, \cite{Vacca2011} measured an M2.5 spectral type for TW Hya in the NIR, while optical temperature measurements placed it as a K7V star \citep{Yang2005}, which implies a $\Delta T \sim 300$~K for TW Hya. \cite{Flores2019} used a $K$-band spectrum to derive a $T_{K}=3640^{+94}_{-92}$ for BP Tau, while optical measurements suggested  $T_{optical}=4050$~K \citep{Schiavon1995,Hartigan1995,Grankin2016}, hence a $\Delta T \sim 400$~K is found for BP Tau.  \cite{Gully-Santiago2017} measured a NIR temperature of $T_{NIR}\sim3180-3330$~K for the weak-lined T Tauri star LkCa~4, while several previous studies measured optical temperatures of $T_{optical}\sim4000$~K \citep{Strom1994,Kenyon1995,Grankin2013} implying a $\Delta T \sim 700-800$~K. 

The fact that the IR temperature of T~Tau~N is cooler than its optical counterpart might suggest that the flux observed at different wavelengths is predominantly emitted by physically distinct parts on the surface of the star. One possibility is that the optical temperature mainly emerges from the quiet photosphere of T~Tau~N, while the NIR temperature is more a combination of both cool spots and quiet photosphere weighted by the filling factor of the spots. Such interpretations have been proposed for other young stars including LkCa~4 \citep{Gully-Santiago2017} and V410 Tau \citep{Rice2011}. Invoking cool spots on the surface of young stars is a reasonable argument as young stars have kilo-Gauss magnetic fields which likely create giant starspots on their surfaces. Cool spots on the surface of stars are not limited to low-mass young stars as similar results have also been found for RS CVn, dwarfs, and giant stars, and all these cases are thought to be related to magnetic activity \citep[see for example][]{Berdyugina2005,Strassmeier2009}.

\cite{Berdyugina2005} showed that the temperature difference between the quiet photosphere and spotted regions appears to be larger for hotter stars than for cooler stars, with values of $\Delta T \sim2000$~K for late F and early G stars, and only of $\Delta T \sim200$~K for the late M stars. \cite{Catalano2002} used line-depth ratios to derive spotted vs. quiet photosphere temperature ratios of $T_{spot}/T_{phot} = 0.8$ for the active RS~CVn stars VY~Ari, IM~Peg, and HK~Lac. \cite{Frasca2008} found a similar result for the II~Peg and $\lambda$~And stars. Although the above mentioned results refer to actual spot temperature measurements and not just NIR vs. optical temperatures, it seems noteworthy that the earlier type star T~Tau~N also has a larger temperature difference than the late K and M stars BP Tau and TW Hya.

An alternative explanation for the temperature discrepancy of T~Tau~N is that the optical temperature measured for T~Tau is affected by emission from hot spots on its surface. Hot spots on the surface of young stars should be a common phenomenon as these stars channel material from their circumstellar disk onto the stellar surface through magnetic field lines \citep[e.g.,][]{Romanova2004}. This material travels almost at free-fall velocity and hits the stellar surface creating shocked regions with temperatures of several thousand Kelvin hotter than the quiet photosphere \citep{Bouvier1995}. Due to the $T_{eff}^4$ dependency of the stellar luminosity, any sufficiently large area with a temperature of a few thousand degrees above the photospheric value would dominate the flux, therefore becoming an important contribution to optical observations. $\rm H \alpha$ emission in the spectrum of T~Tau demonstrates that it is an actively accreting source \citep[see for example,][]{White2004}. These findings are further supported by our $K$-band data where we have detected  the $\rm Br \gamma$ line in emission. These multi-wavelength accretion signatures, in addition to the almost pole-on viewing angle toward T~Tau~N (see Appendix \ref{section:appendix_stellar_params}), open the possibility that hot spots could indeed be responsible for the temperature differences observed on the surface of T~Tau~N.

It is beyond the scope of this paper to identify which of these scenarios, or a combination of both, could explain the temperature difference between optical and NIR observations for T~Tau~N. To answer such a question one would require multi-wavelength and preferably simultaneous observations of T~Tau~N accompanied by a sophisticated model to account for more than two temperature components on the surface of the young star. In this work, we limit ourselves to point out that 1) optical and NIR temperatures are drastically different with a temperature difference of $\Delta T \sim 1000$~K, 2) when a simple two component temperature model is used to match the observed temperatures, an extremely large cold spot coverage is required ($\geq 80 \%$), possibly suggesting that a third temperature component is needed (see Appendix \ref{sec:appendix_two_component_models}), and 3) the mass of T~Tau~N derived from stellar evolutionary models is likely to be overestimated when only the optical temperature is considered.

\subsection{T Tauri as an ejected protostar}

Non-hierarchical triple systems are dynamically unstable, eventually leading to the ejection of one component into either a distant highly eccentric orbit or into an escape \citep[e.g.,][]{Anosova1986}. The ejected component is often, but not always, the least massive in the system \citep[e.g.,][]{Valtonen1991}. Newborn stars are surrounded by circumstellar material, which becomes severely disturbed in binaries during periastron passages, leading to outflow events that will appear as giant Herbig-Haro flows \citep{Reipurth2000}. Numerical simulations show that about half of all triple systems break up during the protostellar stage, and when a protostellar companion is ejected out of its nascent cloud, it becomes an \textit{orphaned protostar} \citep{Reipurth2010}. 


\begin{figure}[ht!]
\epsscale{1.16}
\plotone{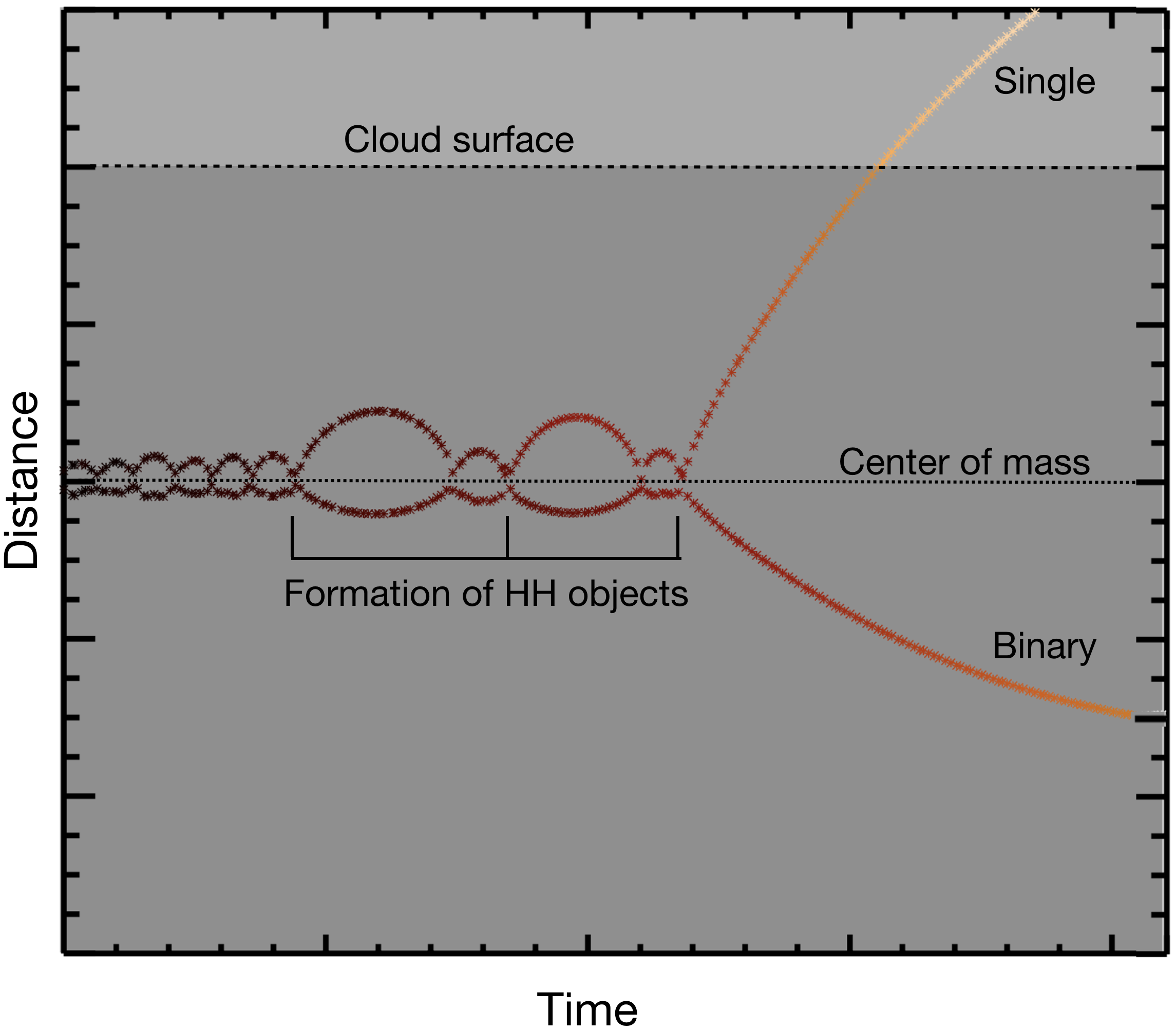}
\label{fig:ejection}
\caption{An illustration of the dynamical processes taking place in newborn unstable triple systems. The plot shows the excursions of the single and binary components following dynamical interactions as a function of time.  One of the three protostars is repeatedly launched from the system following triple ejection events, during which the circumstellar disks interact and drive major accretion/outflow bursts. If born near the surface of the cloud, the ejectee can become optically visible as an orphaned protostar \citep{Reipurth2010}.}
\end{figure}

T~Tau is potentially an example of such a situation. In the above dynamical picture, T~Tau~N has interacted with its siblings leading to several close triple encounters resulting in the large HH~355 Herbig-Haro bow shocks. Assuming a typical tangential velocity for HH objects of 100~$\rm km \,s^{-1}$, a dynamical age of a few thousand years is suggested for HH~355 \citep{Reipurth1997}. It is conceivable that during the last of these encounters T~Tau~N was ejected, and given its projected separation from T~Tau~S of about 100~au, it follows that its mean tangential velocity relative to the T~Tau~S system is approximately 0.2~$\rm km \,s^{-1}$. This is far less than the escape velocity from T~Tau~S, indeed T~Tau~N would have to be ejected almost straight towards the observer for its deprojected velocity to exceed the escape velocity from T~Tau~S. Therefore it is almost certain that T~Tau~N is in a bound orbit around T~Tau~S. We conclude that T~Tau~N is currently breaching the surface of the dense globule in which it was born, and will soon be moving back into the cloud (see Figure \ref{fig:ejection}). A very similar situation, but on a smaller scale, is taking place with T~Tau~Sb, for which a good orbit exists. \cite{Kohler2020} show how Sb was very faint when it was discovered, then gradually brightened as it moved away from Sa, peaking in brightness when it was most distant from Sa. In recent years Sb has been fading again as it approaches Sa, diving into a compact dusty disk, which was presumably the birth site of all three components.

In view of the above, it is thus not surprising that T~Tau~N has a low gravity and a high luminosity: it is an orphaned protostar, still high up on its Hayashi track, and it has been briefly ejected out of its nascent cloud. As T~Tau~N approaches the T~Tau~S binary it will again become part of a non-hierarchical triple system, with the ensuing chaotic motions. In the future, it is therefore likely that one of the three T~Tau components will be ejected into an escape, leaving behind another young binary. Statistical studies of young binaries show that such breakup is common \citep{Reipurth2010} and observations reveal that it occurs primarily at very early evolutionary stages \citep{Connelley2008a,Connelley2008b}.

In the title of this paper, we posed a question that is partly physical and partly semantic. T Tauri stars are divided into Classical TTS and weakline TTS, depending on whether the H$\alpha$ emission is stronger or weaker than a 10~\AA\ equivalent width \citep{Herbig1988}. So from a purely classification point of view, T~Tau~N is clearly a classical T~Tauri star, as expected from an archetypal object. But in terms of its evolution, it is not a typical T~Tauri star. We have argued above that stellar dynamics has played a major role in the history of T~Tau~N, and that it is really a protostellar object that has lost much of its extended envelope by being dynamically ejected out of its birth environment. But what has not changed is that the T~Tau~N star itself is still a highly luminous, bloated star that remains high up on its Hayashi track. So from an evolutionary point of view, it would be incorrect to label it as a typical classical T~Tauri star, when in reality it is an orphaned protostar. So the answer to the question in the title is... `It depends.'

\section{Conclusions}
\label{sec:Conclusion}

We obtained $K$- and $H$-band spectra of T~Tau~N with iSHELL. We modeled the $K$-band spectrum using the MoogStokes magnetic radiative transfer code and derived stellar parameters for T~Tau~N. Furthermore, we used the $H$-band spectrum and an EW line-depth relations method to confirm the $K$-band temperature of T~Tau~N. A summary of our findings are as follows:

\begin{itemize}
    \item Synthetic stellar models were used to measure a temperature of $T_K = 3976^{+90}_{-90}$~K in the $K$-band of T~Tau~N. Empirical temperature relations in the $H$-band verified this result, from which a $H$-band temperature of $T_H = 4085 \pm 155$~K was derived. The surface gravity of this star was measured for the first time yielding a value of $\log{g} = 3.45^{+0.14}_{-0.14}$. Additionally a magnetic field strength of $\langle B \rangle = 1.99^{+0.04}_{-0.05}$~kG, a projected rotational velocity of $v\sin{i}=20.8^{+0.17}_{-0.13}$, and an IR $K$-band veiling value of $r_K = 3.0^{+0.04}_{-0.04}$ were measured, which are all consistent with previous literature results. The CO region in the $K$-band spectrum of T~Tau~N was excluded when deriving stellar parameters for the star (except for $v\sin{i}$), as it showed clear signs of disk emission.
    \item We compared the temperature and gravity values derived for T~Tau~N with a sample of 24 CTTS from Taurus and found that T~Tau~N is among the lowest gravity and hottest objects in the sample. Furthermore, when comparing these results with magnetic evolutionary models from \cite{Feiden2016}, we found that T~Tau~N is the most luminous and youngest star of the sample. Although we could not derive a stellar mass for T~Tau~N, our IR temperature measurements suggest that previous stellar mass determinations from optical studies may have overestimated its value.
    \item We are confident that the temperature difference of $\Delta \rm T\sim1000$~K between optical and NIR results found for T~Tau~N is real. This is confirmed by numerous optical spectroscopic measurements performed on this star, in addition to our two temperature measurements in the $H$- and $K$-bands. Although we cannot uniquely explain this temperature difference, we suggest that a mixture of cool and hot spots on the surface of T~Tau~N produce significant thermal inhomogeneities. In fact, when a simple two temperature model is used to reproduce the observed optical and IR temperatures, an unreasonably large cold spot coverage is required,  suggesting that a three component temperature model is needed.
    \item The above results establish that T~Tau~N is not a typical T~Tauri star, but rather show clear signatures of being a much younger and luminous protostar. This is not surprising given that it is situated only 0.7 arcsec from the embedded Class~I protostellar binary T~Tau~Sa. We interpret this as evidence of prior dynamical interactions in the T~Tau triple system that ejected T~Tau~N so it is now piercing the surface of the small globule in which the system resides. This accounts for the very bright and variable reflection nebula that surrounds T~Tau~N. Based on statistical arguments, it appears that T~Tau has been ejected into a bound orbit, implying that it will again at some point dive back into the cloud and engage in further dynamical interactions, most likely resulting in the final breakup of the triple system.
\end{itemize}

%

\vspace{5mm}
\newpage

\acknowledgments
We thank the referee for an insightful review which helped to improve this paper. We are grateful to Greg Doppmann for a critical reading of an early version of the manuscript, to Ricardo L\'opez-Valdivia for providing valuable information about the uncertainties in the $H$-band temperature determination method, and to Karina Malgue for helping with the proofreading and editing of the manuscript.  We acknowledge the support of the NASA Infrared Telescope Facility, which is operated by the University of Hawaii under contract NNH14CK55B with the National Aeronautics and Space Administration, and we are grateful for the professional assistance in obtaining the observations from the telescope operators of the IRTF. We are particularly grateful to the iSHELL team for all the help provided during the data reduction process. This research has made use of the SIMBAD database, operated at CDS, Strasbourg, France, and NASA’s Astrophysics Data System. This work has made use of the VALD database, operated at Uppsala University, the Institute of Astronomy RAS in Moscow, and the University of Vienna. The technical support and advanced computing resources from the University of Hawaii Information Technology Services – Cyberinfrastructure are gratefully acknowledged.

\facilities{IRTF}


\software{astropy \citep{2013A&A...558A..33A}
          MoogStokes \citep{Deen2013}, 
          emcee \citep{Foreman-Mackey2013}, 
          Spextools \citep{Cushing2004}, 
          xtellcor \citep{vacca2003}
          }



\bibliography{sample63}{}
\bibliographystyle{aasjournal}

\newpage
\appendix

\section{Modification of the VALD3 Line Transition Parameters}
\label{section:vald3_modification}
\begin{deluxetable*}{lcccccc}
\tabletypesize{\scriptsize}
\tablecaption{VALD3 Line Parameters Modified in This Work. \label{table:Line_params}}
\tablewidth{0pt}
\tablecolumns{7}
\tablehead{ \colhead{} & \colhead{} &\colhead{Default VALD3 parameters} & \colhead{} & \colhead{} & \colhead{Modified VALD3 parameters} & \colhead{}\\
\colhead{Element} &\colhead{Wavelength (\AA)} & \colhead{log(gf)} & \colhead{Waals} & \colhead{Wavelength (\AA)} & \colhead{log(gf)} & \colhead{Waals}}
\startdata
 Ti I & 21788.890 & -1.170 & -7.790 & 21788.890 & -1.260  &  -7.590\\
 Ti I & 21903.368 & -1.470 & -7.790 & 21903.348 & -1.500  &  -7.390 \\
\enddata
\end{deluxetable*}

We followed the same treatment as described in \cite{Flores2019} to modify the VALD3 line transition parameters of two Ti lines in the $K$-band spectrum. Namely, we first computed a synthetic solar nonmagnetic spectrum using MoogStokes with standard solar parameters $T_{eff}= 5778$~K, $\log{g}=4.43$, metallicity $\rm \left[ M/H \right]=0.0$, $v_{micro}=1.0$~$\rm km\, s^{-1}$, and $v\sin{i} = 2.0$~$\rm km\, s^{-1}$ and then compared it to solar observations obtained by observing the Sun in reflected light from the asteroid Ceres.

Although the mismatch between the observations and the models with standard VALD3 values could be caused by, but are not limited to, inaccurate $\log{gf}$ and van der Waals constant values in the database, by oversimplifications made in the stellar atmospheric models (such as the LTE assumption), by imperfections in the radiative transfer code, or by an imperfect characterization of our instrumental spectral profile, we assumed that the main source of discrepancy comes from the line transition parameters in the VALD3 database. For this reason, we adjusted the $\log{gf}$ and  van der Waals values of the regions 3 and 4 of Table \ref{table:wavelength_regions} until the computed synthetic spectrum matched the solar observations. All the other regions were already modified in \cite{Flores2019}. 
A summary of the modified line transition parameters can be found in Table \ref{table:Line_params}.

\section{Stellar Parameters from Stellar Inclination}
\label{section:appendix_stellar_params}
It is possible to directly infer the stellar radius of a star when the projected rotational velocity, the stellar rotational period, and the stellar inclination angle are known (see Equation \ref{eq:radii_and_inclination}). In Section \ref{sec:Results}, we have derived a projected rotational velocity of $v\sin{i}=20.8 \,\rm km \, s^{-1}$ for T~Tau~N, and literature values of its rotation period are consistently $\rm P=2.81$~d \citep{Artemenko2012,Rebull2020}, however, no direct stellar inclination measurements exist for this star, so only $R\sin{i}$ can be determined:

\begin{equation}
\label{eq:radii_and_inclination}
    R\sin{i} = \frac{v \sin{i}}{2\pi} P
\end{equation}

Measurements of stellar inclinations often require intense observations and modeling work. This is the case, for example, for inclinations derived with the Doppler Imaging technique from a series of spectroscopic observations \citep[see for example][]{Hill2019,Donati2019}. Unfortunately, this type of observations have not been reported for T~Tau~N. 

Less reliable stellar inclination angles can be derived by observing the circumstellar material around the star. High contrast NIR imaging and coronographic observations of the T~Tauri system revealed that the disk around T~Tau~N has a relatively face-on inclination \citep{Kasper2016}, and recent 1.3 millimeter ALMA  observations of the T~Tauri system showed that the inclination of the dusty disk around T~Tau~N is $i=28.2\degr\pm0.2$ \citep{Long2019}. The later results, however, mostly trace the outer parts of the disk and might not provide a good estimate of the stellar inclination angle. Other studies have also attempted to derive the stellar inclination angle for T~Tau~N in the past. For example, \cite{Bouvier1995} combined optical spectral type and stellar luminosity  measurements to estimate the stellar radius of T~Tau, from which they inferred an equatorial rotational velocity and used it to derive an inclination angle of $i=15\degr$ for this star.

\begin{figure}[ht]
\epsscale{1.2}
\plotone{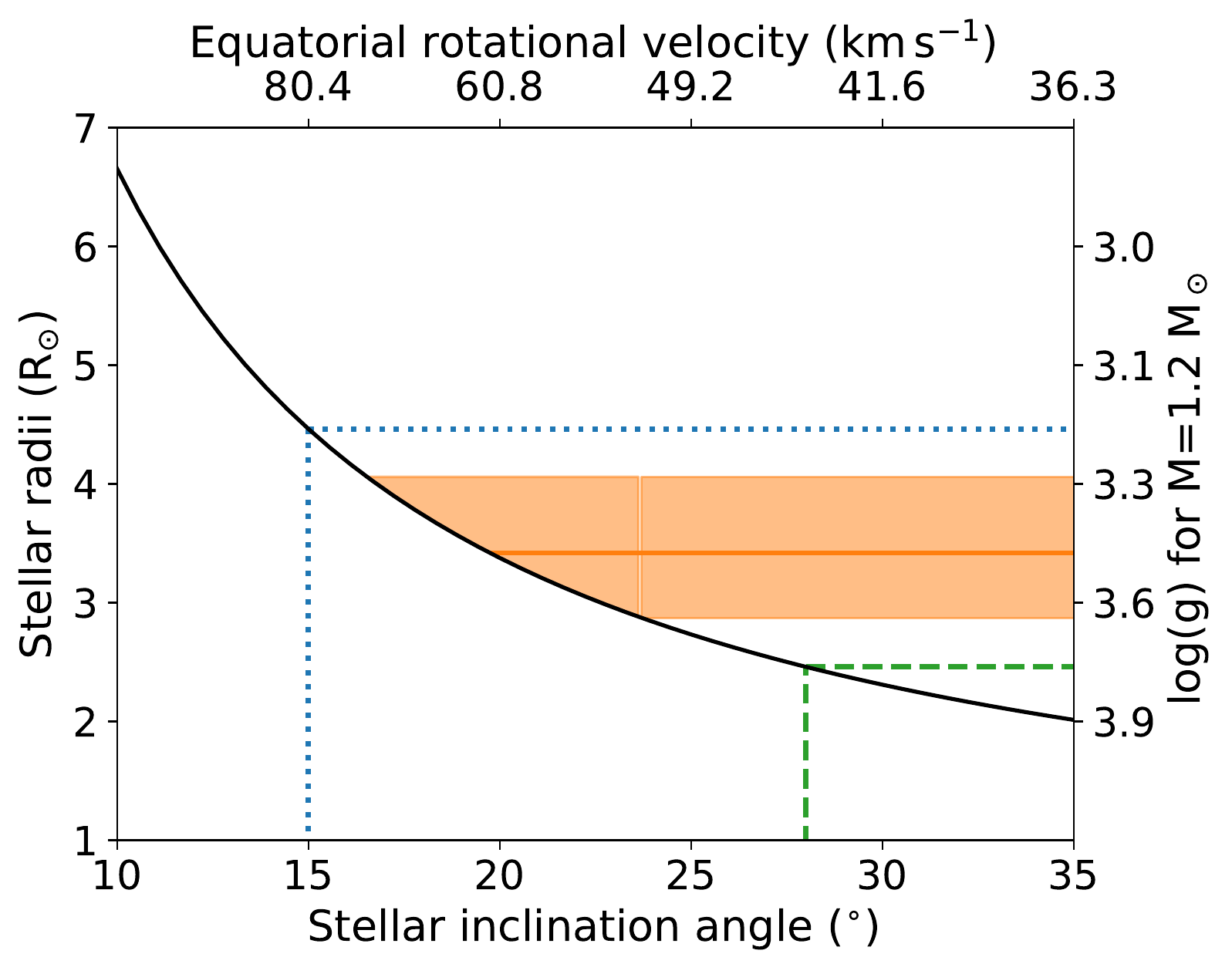}
\caption{Predicted stellar radii as a function of stellar inclination angle assuming a rotation period of $\rm P= 2.81$~d and a projected rotational velocity of $v\sin{i}=20.8 \, \rm km \, s^{-1}$ (black solid line). The green dashed line corresponds to the inclination angle measured for the dust disk around T~Tau~N. The blue dotted line represents the stellar inclination angle derived from \cite{Bouvier1995} (see text). The top axis shows the corresponding equatorial velocity of T~Tau~N. The right axis displays the surface gravity of the star for an assumed stellar mass of $\rm M = 1.2 \, M_\odot$. The orange line and orange shaded region correspond to our gravity measurement and its corresponding uncertainty.\label{fig:radii_from_inclination}}
\end{figure}

In Figure \ref{fig:radii_from_inclination}, we plotted the stellar radius as a function of the stellar inclination angle using the projected rotational velocity and rotation period derived for T~Tau~N. We show the disk's inclination angle of $i=28.2\degr$ measured with ALMA \citep{Long2019} as a green dashed line, and  the inclination angle inferred from the optical spectral type and stellar luminosity by \cite{Bouvier1995}, as a blue dotted line. These two measurements predict vastly different stellar radii of 2.46 $R_\odot$ and 4.46 $R_\odot$, respectively, as well as different equatorial rotational velocities for T~Tau~N (top axis).

Our gravity measurement of $\log{g}=3.45\pm0.14$ for T~Tau~N lies right in between the inferred gravity for the two different inclination measurements mentioned above. Although we assumed a stellar mass of $\rm M = 1.2 \, M_\odot$, we point out that changes in this value do not substantially alter the values on the right axis. For example, if we assume stellar masses of $1.5 \, M_\odot$ and $2.0\, M_\odot$ instead, the corresponding inferred gravities from the green line are $\log{g}=3.83$ and $\log{g}=3.95$, and for the blue line are $\log{g}=3.31$ and $\log{g}=3.44$.

\section{A two component temperature model for T~Tau~N}
\label{sec:appendix_two_component_models}

The questions we aim to answer in this section are the following: (1) if we use a two component model with a hot component ($T_{hot}$), a cold component ($T_{cold}$), and a geometrical filling factor ($f$), what combinations of these three parameters can reproduce the observed optical $T_{R-band}\sim 5000$~K \citep{Herczeg2014} and K-band  $T_{K-band}\sim4000$~K (see Section \ref{sec:Results}) temperatures for T~Tau~N? and (2), can these parameters provide a realistic description of the surface of the star?

The assumptions made in this two-component model are {\em a}) the cold and hot components are represented by Planck functions $\rm B_{\lambda}(T)$; {\em b}) the filling factor is the geometrical filling factor, i.e., the relative area covered by the cold component as compared to the hot component (see equations \ref{eq:filling_factors_hot} and \ref{eq:filling_factors_cold}); and {\em c}) at each wavelength the observed temperature is the weighted average of the hot and cold temperatures (equation \ref{eq:model_temperature}). In other words, if half the light is from the hot component and half is from the cold component, then the observed temperature from the spectrum would be half way in between them. This assumption is most accurate when the hot and cold temperatures are close to each other.

\begin{eqnarray}
\label{eq:filling_factors_hot}
b_{\lambda , hot} &=& \rm B_{\lambda}(T_{hot})(1-f)  \\
\label{eq:filling_factors_cold}
b_{\lambda , cold} &=& \rm B_{\lambda}(T_{cold})f
\end{eqnarray}

\begin{eqnarray}
\label{eq:model_temperature}
T_{\lambda} &=& T_{hot}  \left(  \frac{b_{\lambda , hot}}{b_{\lambda , hot} + b_{\lambda , cold}} \right) \nonumber \\
&+& T_{cold} \left(  \frac{b_{\lambda , cold}}{b_{\lambda , hot} + b_{\lambda , cold}} \right)
\end{eqnarray}

\begin{figure}[ht]
\epsscale{1.1}
\plotone{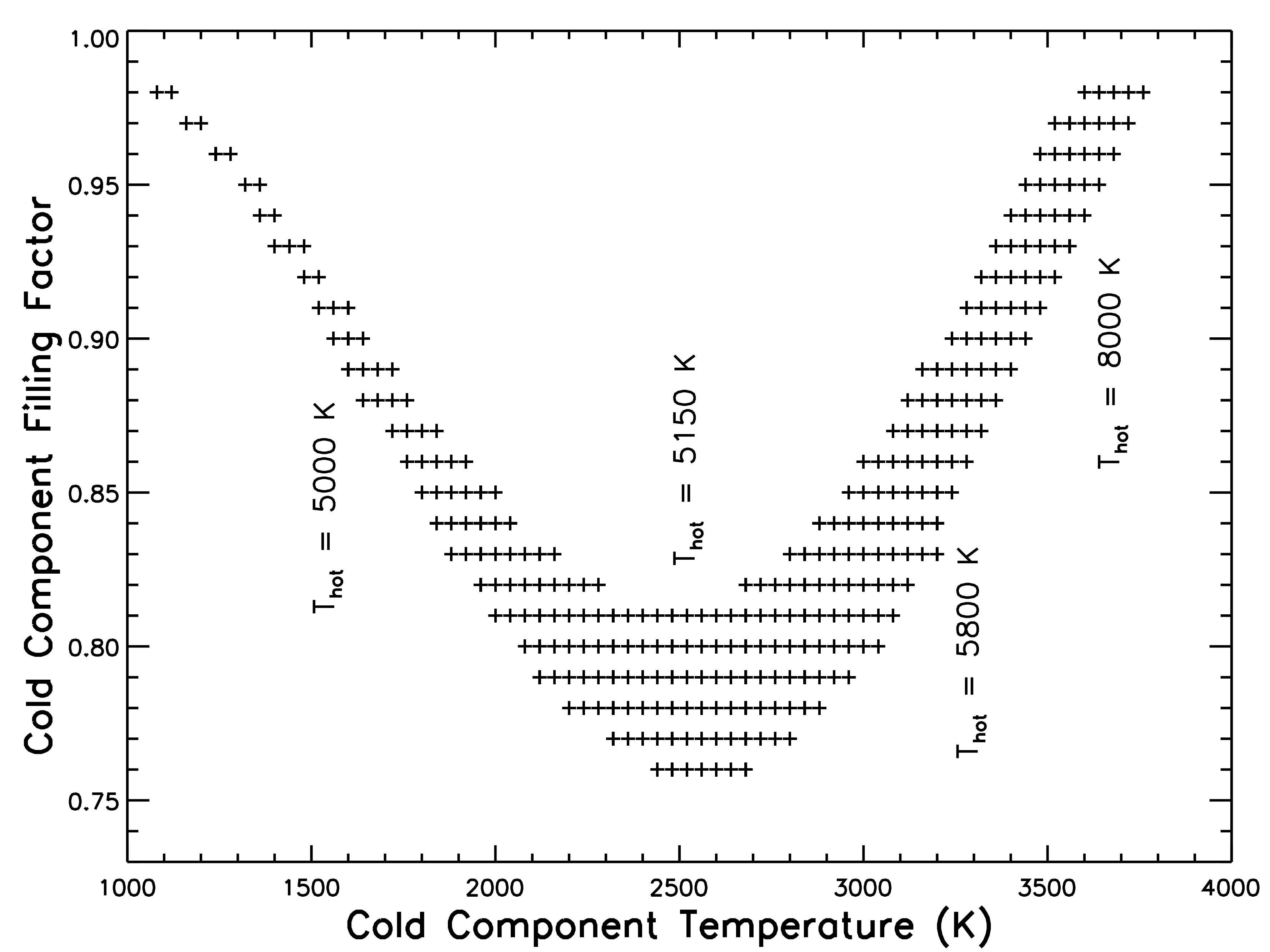}
\caption{Cold component filling factor as a function of the temperature of the cold component. The temperature of the hot component is also shown on the plot and it varies between $\sim5000$~K to $\sim8000$~K. We find that the filling factor of the cold component is always large with a surface coverage of at least 75$\%$, suggesting that a third hot component is also needed.}
\label{fig:two-component-model}
\end{figure}

In Figure \ref{fig:two-component-model}, we show the values of the different components when the calculated optical and infrared temperatures approach the measured $T_{R-band}\sim 5000$~K and $T_{K-band}\sim4000$~K, at their respective wavelengths. Overall we find that the observations can be well matched by a wide range of parameters, but they follow a trend.  If the cold component is very cold ($T_{cold}\leq1300$~K), then the hot component is $T_{hot}\sim5000$~K, and the filling factor must be $f \sim95\%$.  This makes sense if the hot temperature equals the optical temperature and the cold temperature is so cold that it contributes nothing in the optical. If the cold component is $T_{cold}\sim3700$~K then the hot component must be $T_{hot}\sim8000$~K and again the filling factor becomes very large.  Here, the code is making the cold temperature equal the IR temperature and the filling factor is so low that the hot component contributes little to the optical temperature. Naturally, it is not very realistic that adding a 8000~K and a 3700~K spectrum would look like a 4000K spectrum at $K$-band since the hot component would produce the broad hydrogen lines from an A-star. Finally, if the cold component is  $T_{cold}\sim2500$~K, then the hot component is about $T_{hot}\sim5200$~K and the filling factor is roughly $f \sim80\%$.

Although this last model provides the most reasonable hot and cold temperature components, it still requires a very large starspot filling factor. Whether or not a filling factor of the order of $\sim80\%$ is physically realistic is still a matter of debate as spectropolarimetric observations have shown spot filling factors of only 15 to $25\%$, while other techniques such as broad wavelength coverage two-component spectral fitting or even TiO line analysis have achieved filling factors of $\sim80\%$ \citep[e.g., for LkCa4][]{Gully-Santiago2017} and $>50\%$ \citep[e.g., for the RS CVn star II Pegasi][]{Berdyugina2005,Neff1995}, respectively.

We conclude by stating that a three component model would be better suited for an accreting star such as T~Tau~N, and therefore our two component model is unlikely to encompass the full complexity suspected for a young accreting and magnetically active T~Tau~N.




\newpage




\end{document}